\documentstyle[12pt]{article}
\begin{document}
 \baselineskip=13pt
 \setlength{\unitlength}{1cm}
 \setcounter{page}{0}
 \pagestyle{plain}
 \begin{center}
 {\Large{\bf A Critique on Quantum-No-Deleting Principle}} \\
 \vspace{.5cm}
 {K.\ V.\ Bhagwat, D.\ C.\ Khandekar, S.\ V.\ G.\ Menon,
 R.\ R.\ Puri and D.\ C.\ Sahni.} \\
 \end{center}
 {\bf Abstract:}
 {\small
The  argument  used,  in  a recent letter to {\it Nature}, to arrive at the
`quantum-no-deleting principle' is erroneous. It is pointed  out  here  that
there  may  not  be  anything like such a principle. In any case, the claims
made in the letter are beyond its working premise.} \vspace{.5cm}


The   March   issue   of  {\it  Nature}  contains  a  letter  entitled  {\it
`Impossibility of deleting an unknown quantum state}' by A.\  K.\  Pati  and
S.\  Braunstein  [1] (hereafter referred as PB for short). The letter begins
with a reference to an earlier letter by W.\ K.\ Wootters and W.\ H.\  Zurek
[2],  according  to  which  an  unknown  quantum  state  cannot be cloned or
duplicated. (An unknown  quantum  state  is  a  linear  superposition,  with
unknown  coefficients, of the preferred states of a system. For example, the
up and down states, with respect to a specified direction, are the preferred
states  of  a  spin-1/2  particle.)  Then,  the  authors  talk   about
desirability  of  having  an  option  to  delete  information  in  a quantum
computer. They discard the usual method of irreversibly erasing  information
as  of  no  interest,  and  that  they  are  concerned  only with the act of
`uncopying', which is the opposite of copying. Uncopying needs at least  two
copies  of  a  state; the quantum-deleting machine performing uncopying will
operate on two identical input states - the original and a copy - and induce
the copy to switch to some prescribed state. The  reason  for  insisting  on
uncopying,  rather  than  the  usual  erasing,  is very vague in the letter,
except for a reference to Landauer, who had pointed  out  that  irreversible
erasure of information leads to increase of entropy.

In the main body of the letter, PB talk about a transformation, which is the
quantum  mechanical time evolution of a composite system consisting of three
sub-systems. There are two identical sub-systems in identical  states,  each
denoted  by  $\Psi$,  and an ancilla, which represents the remaining part of
the composite system, in a prescribed state $A$.  As  defined  by  PB,  this
system would perform the function of an uncopying machine, if it starts from
a  composite  state,  denoted by $\Psi \Psi A$, and ends up in a state $\Psi
\Sigma A_ \psi$. Thus, uncopying is to switch the second $\Psi$ to  $\Sigma$
,  which  is  a prescribed state of the second sub-system. The authors agree
here that, in this process, the state $A$  of  the  ancilla  is  changed  to
$A_{\Psi}$.  The  explicit  sub-script $\Psi$ on $A$ is to indicate that the
final state of the ancilla depends on the state to be uncopied. The  authors
assume  that  such  a transformation exists when $\Psi$ is $H$ or $V$, which
are  the  preferred  polarization  states   -   horizontal   and   vertical,
respectively - of a photon. Thus, it is hypothesized that under the transformation
(i) $H H A$ $\rightarrow$ $H \Sigma A_H$ and (ii) $V V A$ $\rightarrow$ $V \Sigma A_V$, where $A_{H}$
and  $A_{V}$  are,  respectively,  the  states  of  the  ancilla  after  the
transformations. Then, their question is whether the same transformation can
uncopy an arbitrary linear superposition, $\Psi  =  \alpha  H  +  \beta  V$,
normalized  to  unity.  After  some elementary steps, they find that this is
indeed possible with an appropriate $A_{\Psi}$. The composite state, $( H  V
+  V  H  )  A$,  then  transforms  to $(H \Sigma A_{V}+V \Sigma A_{H})$. The
required form for $A_{\Psi}$ is $(\alpha A_{H}+  \beta  A_{V})$.  Thus,  the
conclusion  should  have  been  that  the sought after transformation on the
composite state vector is indeed possible - even  though  it  has  not  been
explicitly  constructed.
\par   However,   at   this  point PB change their attitude. They
argue that since $A_H$  and $A_V$ are orthogonal, their equation  (1)
implies  {\it swapping}  and not uncopying. They then announce  the  birth
of a new `quantum-no-deleting principle' in the letter.
\par While it is ture  that in  the
case of swapping  of  the  states $H$ and  $V$  (with  the  state
$\Sigma$  from  the  ancilla) the  corresponding ancilla   states
$A_H$  and $A_V$
are {\it indeed} orthogonal, the converse,  tacitly used by PB, is
however,   not true.  This is  demonstrated     by   the following
counter-example. Let  the initial state of the ancilla be
$A  = \Psi_1 \Psi  _2 ...  \Sigma \Psi_l ...$
and  let  the  evolution be such  that the  $A_H$  and $A_V$ of
transformations  (i)  and (ii) are given  by
$A_H  = \Psi_1 \Psi  _2 ...[ {1 \over \sqrt{2}} (H+V)] \Psi_l ...$ and
$A_V  = \Psi_1 \Psi  _2 ...[ {1 \over \sqrt{2}} (H-V)] \Psi_l ...$.
Obviously,  the   evolution  does   not represent  swapping. But
$(A_H,A_V) =0$. This  counter-example clearly shows   that  vanishing  of
the  scalar  product   of  $A_H$ and $A_V$ {\it does  not  imply}
that the underlying  evolution represents swapping.

Many more  claims  are  sprinkled  throughout  the  letter:  e.g.;  (i)  the
no-deleting principle is claimed to have been derived for an unknown quantum
state   (the  unknown  character  is  never  used  at any  stage  of  their
`derivation'), (ii) the no-deleting principle is claimed to be  deduced  for
reversible  as  well as irreversible machines ( although  Schr\"{o}dinger
evolution, that is always reversible, is tacitly assumed by PB), etc.

Towards  the  last  paragraph  of  the letter, PB even forget that they were
considering only the narrow act of uncopying, and go on to make still bigger
claims. {\it $"$We emphasize that copying and deleting of information  in  a
classical  computer  are  inevitable  operations  whereas similar operations
cannot be realized perfectly in quantum computers. This may  have  potential
applications   in  information  processing  because  it  provides  intrinsic
security to quantum files in a quantum computer. No  one  can  obliterate  a
copy  of  an  unknown  file from a collection of several copies in a quantum
computer. In spite of the quantum no-deleting principle, one  might  try  to
construct  a  universal and optimal approximate quantum- deleting machine by
analogy with quantum cloning machines. When memory in a quantum computer  is
scarce  (at  least  for a finite number of q-bits), approximate deleting may
play an important role in its own way.  Although  at  first  glance  quantum
deleting  may seem the reverse of quantum cloning, it is not so. Despite the
distinction between these two operations, there may be some link between the
optimal fidelities of approximate deleting and cloning. Nevertheless, nature
seems to put another  limitation  on  quantum  information  imposed  by  the
linearity  of quantum mechanics$"$.} All these very high-sounding claims lie
well outside the premises of the matter of their discussion.

In  the  same  issue  of  {\it Nature}, W. H. Zurek [3], has expressesed his
views on quantum cloning with a write-up {\it $"$Schr\"{o}dingers sheep$"$},
and has referred to the letter of PB. He says that PB's result complements
his no-cloning theorem. How do we understand this remark in the light of the
above discussion? Zurek explains, in some detail, how the  impossibility  of
reversing   a   sequence   of   logical   steps   leads   to   thermodynamic
irreversibility, and shows how a (classical) logic operation, called  C-NOT,
can  reversibly  delete  a  (classical) bit against a copy. This logic gate,
which can keep on functioning without increasing entropy, accepts two inputs
- the original and a copy - it does nothing if the original bit  is  0,  but
flips  the copy if the original bit is 1. Zurek says that a quantum C-NOT is
also a physically realizable  system  -  it  could  be  a  composite  system
consisting  of  two spin-1/2 particles - evolving as per the Schr\"{o}dinger
equation. Thus, like its classical counterpart, the states ${\underline  0}\
{\underline 0}$ and ${\underline 0}\ {\underline 1}$ will evolve to the same
states,  while  the states ${\underline 1}\ {\underline 0}$ and ${\underline
1}\ {\underline 1}$ will end up as  ${\underline  1}\  {\underline  1}$  and
${\underline  1}\  {\underline  0}$.  This  system  can  perform copying and
deleting operations, on these preferred states, just like in  the  classical
case.  However,  Zurek  shows  that  both operations are not possible if the
starting state is ${\underline S}\ {\underline S}$, where  ${\underline  S}=
\alpha  {\underline  0}+  \beta  {\underline 1}$, a linear superposed state.
This is so irrespective of the fact whether ${\underline  S}$  is  known  or
unknown,  as  C-NOT  does  not  use $\alpha$ and $\beta$ at all. The unitary
property of C-NOT ensures that  if  it  cannot  transform  ${\underline  S}\
{\underline 0}$ to ${\underline S}\ {\underline S}$, (i.e. copying), then it
cannot take ${\underline S}\ {\underline S}$ to ${\underline S}\ {\underline
0}$ (i.e. uncopying) either. We believe that this is, precisely, the meaning
of Zurek's remark. (He has also cautioned against neglect of de-coherence of
quantum  correlation.  Otherwise, one can end up in paradoxes.) Furthermore,
Zurek argues that C-NOT may be modified by  adding  more  components  to  it
(MC-NOT  say).  He  then  asserts  that  with  MC-NOT  cloning  or  deleting
superposed states, with known values of $\alpha$ and $\beta$, is not at  all
a  problem.  MC-NOT  would  accept values of $\alpha$ and $\beta$ as inputs,
accordingly rotate the vector ${\underline S}$ to either ${\underline 0}$ or
${\underline 1}$, perform the needed operation, and finally, rotate it  back
to   ${\underline   S}$.  If  ${\underline  S}$ is unknown, using
MC-NOT in place of
C-NOT for cloning will not do any better. Thus, no-cloning theorem for  {\it
unknown  quantum  states}  is  inescapable.  The  situation for uncopying as
defined by PB seems different. As pointed out above, it is possible to  have
a quantum evolution for a composite system, consisting of three sub-systems,
which will do uncopying as defined by PB; since the third sub-system is free
to  adjust  itself.  As  their  definition  does  not involve the `known' or
`unknown' character of $\Psi$, uncopying is always possible! Thus,  there is
nothing  like  a quantum-no-deleting principle even within the limited scope
of uncopying.

In an early paper, (which is also referred to by PB), H.\ P.\ Yuen [4] shows
that  in  principle  a device exists which would duplicate a quantum system,
within a class of quantum states, if and only  if  the  quantum  states  are
orthogonal.  This theorem, which is a rigorous expression of all the aspects
of  the  cloning  problem,  is  deduced  using  the  unitary   property   of
Schr\"{o}dinger  evolution.  In  fact,  Yuen  considered  a  three-component
copying device aimed for transforming a composite vector $\Psi \Sigma A$  to
$\Psi  \Psi A_{\Psi}$, where $\Psi$ belongs to a set of two or more linearly
independent states, $\Sigma$, $A$ and $A_{\Psi}$ are as defined earlier.  It
is  clear  that the uncopying machine of PB is simply the inverse of Yuhen's
operator, but with a subtle difference - the status of $A$ and $A_ \psi$ get
interchanged! It is this difference that leads  to  the  feasibility  of  an
uncopying  machine.  If  feasible,  it can uncopy quantum states, `known' as
well as `unknown'!

Finally,  a  word  about the theme of the letter, based on their supposition
would be in order. We recall that in the opening paragraph, PB  write,  {\it
$"$  suppose,  at  our  disposal  we  have  several copies of a photon in an
unknown quantum state$"$}. This is intriguing.
How  can  one  claim identity of two (or several) states that are {\em
completely unknown?} Clearly,  the  theme  of  the  letter  is  based  on
a logically unsound supposition.

\vspace{.5cm}
\noindent {Address for correspondence: \\
Bhabha Atomic Research Centre, \\
Technical Physics and Prototype Engeneering Division, \\
Modular Labs, II Floor, Mumbai, 400 085, INDIA.\\
e-mail: kvb@apsara.barc.ernet.in }
\vspace{.5cm}

\noindent
(1) A.\ K.\ Pati and S.\ Braunstein, {\it Nature}, 164, {\bf 404}, 2000. \\
(2) W.\ K.\ Wootters  and W.\ H.\ Zurek, {\it Nature}, 802, {\bf
299}, 1982.\\
(3) W.\ H.\ Zurek, {\it Nature}, 130,  {\bf 404}, 2000. \\
(4) H.\ P.\ Yuen, {\it Phys. Lett.}, 405, {\bf  113 A}, 1986).
\end{document}